\documentclass[%
 reprint,
 amsmath,amssymb,
 superscriptaddress,
 aps,
]{revtex4-2}

\usepackage{graphicx}
\usepackage{dcolumn}
\usepackage{bm}

\usepackage{mathrsfs}

\renewcommand{\eqref}[1]{\textup{{\normalfont Eq.~(\ref{#1}}\normalfont)}}
\newcommand{\secref}[1]{\textup{{\normalfont Sec.~\ref{#1}}\normalfont}}
\newcommand{\figref}[1]{\textup{{\normalfont Fig.~\ref{#1}}\normalfont}}

\begin{document}

\author{Paul Johanns}
 \affiliation{
 Flexible Structures Laboratory, Institute of Mechanical Engineering, École Polytechnique Fédérale de Lausanne (EPFL), Lausanne, Switzerland
 }

\author{Paul Grandgeorge}
 \affiliation{
 Flexible Structures Laboratory, Institute of Mechanical Engineering, École Polytechnique Fédérale de Lausanne (EPFL), Lausanne, Switzerland
 }
\author{Changyeob Baek}%
\affiliation{ 
Department of Mechanical Engineering, Massachusetts Institute of Technology, Cambridge, MA, USA
}%
 \author{Tomohiko G. Sano}
 \affiliation{
 Flexible Structures Laboratory, Institute of Mechanical Engineering, École Polytechnique Fédérale de Lausanne (EPFL), Lausanne, Switzerland
 }
 
\author{John H. Maddocks}
\affiliation{%
Laboratory for Computation and Visualization in Mathematics and Mechanics, Institute of Mathematics, École Polytechnique Fédérale de Lausanne (EPFL), Lausanne, Switzerland
}%
\author{Pedro M. Reis}
 \email{pedro.reis@epfl.ch}
\affiliation{
 Flexible Structures Laboratory, Institute of Mechanical Engineering, École Polytechnique Fédérale de Lausanne (EPFL), Lausanne, Switzerland
 }%


\title{The shapes of physical trefoil knots}


\begin{abstract}
We perform a compare-and-contrast investigation between the equilibrium shapes of physical and ideal trefoil knots, both in  \textit{closed} and \textit{open} configurations. Ideal knots are purely geometric abstractions for the tightest configuration tied in a perfectly flexible, self-avoiding tube with an inextensible centerline and undeformable cross-sections. Here, we construct physical realizations of tight trefoil knots tied in an elastomeric rod, and use X-ray tomography and 3D finite element simulation for detailed characterization. Specifically, we evaluate the role of elasticity in dictating the physical knot's overall shape, self-contact regions, curvature profile, and cross-section deformation. We compare the shape of our elastic knots to prior computations of the corresponding ideal configurations. Our results on tight physical knots exhibit many similarities to their purely geometric counterparts, but also some striking dissimilarities that we examine in detail. These observations raise the hypothesis that regions of localized elastic deformation, not captured by the geometric models, could act as precursors for the weak spots that compromise the strength of knotted filaments.
\end{abstract}

\maketitle


\section{Introduction}
\label{Introduction}

The open trefoil knot, commonly known as the overhand knot, is the most elemental open knot, forming the basis of many, more complex, and more functional knots. The trefoil knot can be regarded as a building block in bend knots (\textit{e.g.}, fisherman's/English knot)~\cite{Wright1928}, in binding knots (\textit{e.g.}, square or reef, and granny knots)~\cite{Maddocks1987, Daily-Diamond2017, Patil2020}, and in noose knots (\textit{e.g.}, lasso noose, honda knot, and lariat loop)~\cite{Ashley1944}.  The classic overhand knot is also key in suturing procedures (\textit{e.g.}, surgeon's knot)~\cite{Taylor1938, Edlich2008, Dubrana2011, Taylor2014, Chow2015, Alden2017}. Overhand knots can form spontaneously in various natural contexts, across a wide range of length scales, from polymers and DNA strands~\cite{Saitta1999, Arai1999, Rybenkov1993} to the umbilical cord of human fetuses~\cite{Goriely2005}, and even in vortex loops in plasma and fluid flows~\cite{Kleckner2013, Kleckner2016, Cirtain2013}. 

The classic \textit{mathematical theory of knots} is largely concerned with all possible topologies of knots tied in a closed loop. More recently, a smaller mathematical literature of the \textit{geometry} of so-called \textit{ideal} knot shapes has been developed~\cite{Calvo2005}. For context, in Appendix~\ref{Appendix:Ideal_shapes}, we provide a brief review of recent advances on the theory of ideal knots that may be of interest to the Mechanics community. In this purely geometric context, a knot is modeled as being tied in a closed loop of idealized rope approximated as a filament with an undeformable circular cross-section, inextensible centerline, and vanishing bending stiffness. The ideal, or tightest, shape is then the centerline configuration of a closed tube with the given knot type and diameter $D_0$ that has the shortest possible length $L_0$. For example, an \textit{unknot} is any configuration of a closed loop that can be smoothly deformed to a circle without passing through itself. Unsurprisingly, the ideal shape of the unknot is a circle of circumference $L_0=\pi D_0$. Interestingly, the unknot is the only knot for which the ideal shape is known explicitly; all other ideal knot shapes have only been approximated numerically. The trefoil knot is the simplest nontrivial knot type, and numerical approximations are available, computed with a variety of algorithms, with the most accurate shape obtained to date having $L_0/D_0 = 16.371476 \dots$~\cite{Przybyl2014}. Further geometric characteristics of the ideal closed trefoil are given in Appendix~\ref{Appendix:Ideal_cTrefoil}.

Ideal shapes of \textit{open} knots can also be defined, for which both the diameter and a (long) arc length of the filament are prescribed. The ideal shape for a given knot type arises for the configuration with the maximal distance (in space, not arc length) between the two ends of the filament. This is a mathematically well-posed notion that was
first simulated by Piera{\'{n}}ski~\textit{et al.}~\cite{Pieranski2001, Pieranski2001a}. These authors also sought to relate the equilibrium shape of a knotted filament to the decrease in its mechanical strength, as induced by the knot itself. They reported peaks in the curvature profile along the knot at both the entrance and exit points of the knot. Consequently, it was hypothesized that the weakening of knotted filaments, commonly confirmed by practical experience, was rooted in these geometric features. This observation has also been corroborated at the atomic scale by Saitta~\textit{et al.}~\cite{Saitta1999}, who performed molecular dynamic simulations on knotted polymer strands and pointed to a strain-energy localization at the entrance and exit to the open trefoil knot. However, more recent studies by Uehara~\textit{et al.}~\cite{Uehara2007} and Przyby{\l}~\textit{et al.}~\cite{Przybyl2009} suggest that the ideal rope model may not be appropriate to describe the \textit{mechanical} properties of \textit{tight} physical knots. Whereas recent studies have addressed the mechanics of \textit{loose} overhand knots~\cite{Audoly2007, Clauvelin2009, Jawed2015}, the mechanics of the corresponding \textit{tight} configurations remains largely unexplored.

\begin{figure}[ht!]
        \centering
        \includegraphics[width=0.8\columnwidth]{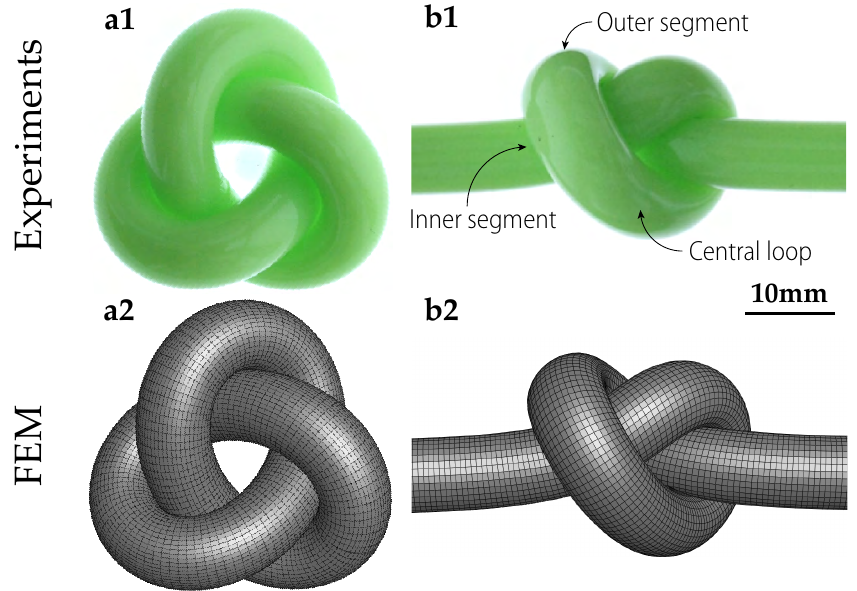}
        \caption{\textbf{Elastic \textit{closed} and \textit{open} trefoil knots in experiment and FEM.} \textbf{a1},~Experimental photograph of a closed trefoil knot tied on an elastomeric rod with length-to-diameter ratio $L_0/D_0 = 16.37$ ($L_0=139.1\,\text{mm}$ and $D_0=8.5\,\text{mm}$). \textbf{a2},~Numerical counterpart of \textbf{a1} computed from FEM. \textbf{b1},~Experimental photograph of a tight, open trefoil knot tied on an elastomeric rod. \textbf{b2},~Numerical counterpart of \textbf{b1} computed from FEM.}
        \label{Fig1}
\end{figure}

Here, we perform a compare-and-contrast investigation between the equilibrium shapes of physical realizations of tight elastic trefoil knots and those of ideal knots based on existing purely geometric models~\cite{Pieranski2001,Carlen2005}, both in \textit{open} and \textit{closed} configurations. We realize physical knots tied onto elastomeric rods (which are straight in their unstressed configuration) in experiment complemented with fully 3D elastic simulation using the finite element method (FEM); representative examples are provided in the experimental photographs and FEM-snapshots of~\figref{Fig1}. Data from X-ray micro-computed tomography ($\mu$CT) are used for a thorough quantitative validation of our FEM computations. Firstly, we focus on the \textit{closed} trefoil knot, given its advantage of having a closed centerline with periodic boundary conditions; in particular, no external forces are required to attain equilibria. In its tight equilibrium configuration, a 2D mapping of the contact \textit{surface} in the physical knot revealed that the double-contact \textit{lines} first computed by Carlen~\textit{et al.}~\cite{Carlen2005} within the purely geometric model form an accurate outer skeleton for the contact surface patch observed in the elastic case. Secondly, we study tight configurations of the \textit{open} trefoil knot, where different levels of tightness can be systematically investigated by the application of a range of external forces, thereby elucidating the effects of elasticity. Our measured curvature profiles for knotted elastic filaments, both in the closed and open trefoils, are qualitatively different from those predicted by the ideal geometric models. Specifically, physical open knots exhibit curvature peaks inside the knot, instead of at their entrance/exit, contrary to previous predictions for the tightest ideal knot~\cite{Pieranski2001}. The excellent FEM-experimental agreement confirms the observed curvature profiles and enables us to extract and map the contact pressure distribution, thereby revealing significant rod constrictions at the entrance and exit of the tight open knot. Finally, we characterize these regions of localized elastic deformation, which we speculate could act as precursors for the weak spots that compromise the strength of knotted filaments.

\section{Physical realization of trefoil knots}
\label{Trefoil_Exp}

We have devised an experimental framework and performed FEM simulations to realize tight physical knots tied on homogeneous, intrinsically straight, elastic rods. In this section, we describe the methodology that we followed on both fronts.

\subsection{Experimental protocols}

\noindent \textit{2.1.1. Fabrication of customized elastic rods:}
We fabricated composite elastomeric rods with the goal of making them compatible with $\mu$CT imaging and 3D image analysis to extract their centerline coordinates and self-contacting regions. 
We used the same fabrication protocol introduced recently to study the contact mechanics between two elastic rods~\cite{Grandgeorge2020}, but with the additional feature described below. The method described in~\cite{Grandgeorge2020} enabled the fabrication of composite elastomeric rods made out of vinyl polysiloxane, VPS32 (Elite Double 32, Zhermack, Young's modulus $E=1.25$~MPa, density~$\rho=1160$~kg/m$^3$), decorated with an elastomeric concentric physical fiber (diameter 500~$\mu$m) and a 150~$\mu$m-thick elastomeric coating. The concentric physical fiber and the coated layer were made of a different, lighter, elastomer (Solaris, Smooth On, $E_\text{solaris}=320$~kPa, $\rho_\text{solaris}=1001$~kg/m$^3$). The 13\% lower density of Solaris with respect to VPS32 allows for the segmentation of the features of interest (centerline and the self-contact regions) during post-processing stages of the $\mu$CT tomographic images, as detailed in Ref.~\cite{Grandgeorge2020}. In the present work, we introduce an additional feature to our composite rods by embedding a second, \textit{eccentric}, physical fiber made of Solaris and diameter 500~$\mu$m, parallel to the concentric fiber, at a distance of 2~mm. This inset fiber allowed us to match the twist of the glued extremities when fabricating the \textit{closed} trefoil knot. Finally, the elastomeric rods of total diameter $D_0 = 8.5\,$mm were then cut to different values of their total length of $L_0$, depending on the system of interest.\\

\noindent \textit{2.1.2. Tying of open trefoil knots:} 
We tied open trefoil knots on the fabricated rods.
Any build-up of excess twist at the free ends was avoided by carefully aligning the eccentric fiber at the extremities during the manual tying process. The knot was progressively tightened by increasing the end-to-end distance while the sample was immersed in a container of soapy water (Palmolive Original) to ensure vanishing friction conditions. The limited size of the sample holders of the $\mu$CT apparatus required rods of different undeformed lengths: $L_0 = 185\,$mm or $125\,$mm, respectively, for the looser or tighter open knot configurations detailed next. Piera{\'{n}}ski~\textit{et al.}~\cite{Pieranski2001} computed the normalized knot length, $\Lambda_\textrm{OC}$ (the engaged knot length divided by the tube diameter), corresponding to the normalized difference between the arc lengths of the centerline associated to the first (entrance) and last (exit) contact points, $s_1$ and $s_2$, respectively. Both the $\mu$CT scanning and the FEM provide access to $\Lambda_\textrm{OC}$.  We chose two elastic configurations, one looser ($L_0=185\,\text{mm}$, $\Lambda_\textrm{OC}=127.5/D_0=15.0$) and the other tighter ($L_0=125\,\text{mm}$, $\Lambda_\textrm{OC}=85.9/D_0=10.1$, see \figref{Fig1}b1) than the tightest ideal open trefoil knot ($\Lambda_\textrm{OC}=12.4$)~\cite{Pieranski2001}.\\

\noindent \textit{2.1.3. Tying of the closed trefoil knot:} 
To compare the elastic closed trefoil knot and its ideal equivalent (results in \secref{cTrefoil_Results}), we trimmed the elastic rod according to the length-to-diameter ratio computed by \textit{e.g.}\ Carlen~\textit{et al.}~\cite{Carlen2005}. For these experiments, our undeformed  elastic rod of diameter $D_0 = 8.5\,\text{mm}$ had a length of $L_0 = 16.37 D_0 \approx 139.1\,\text{mm}$ dictated from the geometric model. The physical closed trefoil knot was tied by first producing an open trefoil knot and then joining the two rod extremities using a silicone adhesive (Sil-Poxy, Smooth-On). 
During this closure procedure the eccentric fibers at each end were aligned at the joint location, which appeared to closely correspond to minimizing any additional, imposed, excess twist. The closed knot was placed in an ultrasonic bath (VWR, USC1200TH) with a water--soap mixture (Palmolive Original, $\approx$20\%
in volume) for five minutes (at frequency $45\,\text{kHz}$ and temperature $22\,^\circ$C).
The combination of the ultrasonic vibrations and lubrication by the soap minimized frictional effects in the regions of self-contact (ensuring the absence of tangential surface stresses there), in agreement with the assumption of frictionless self-contact of idealized rods. \figref{Fig1}a1 shows an optical photograph of the final physical closed trefoil knot. \\

\begin{figure}[ht!]
        \centering
        \includegraphics[width=0.8\columnwidth]{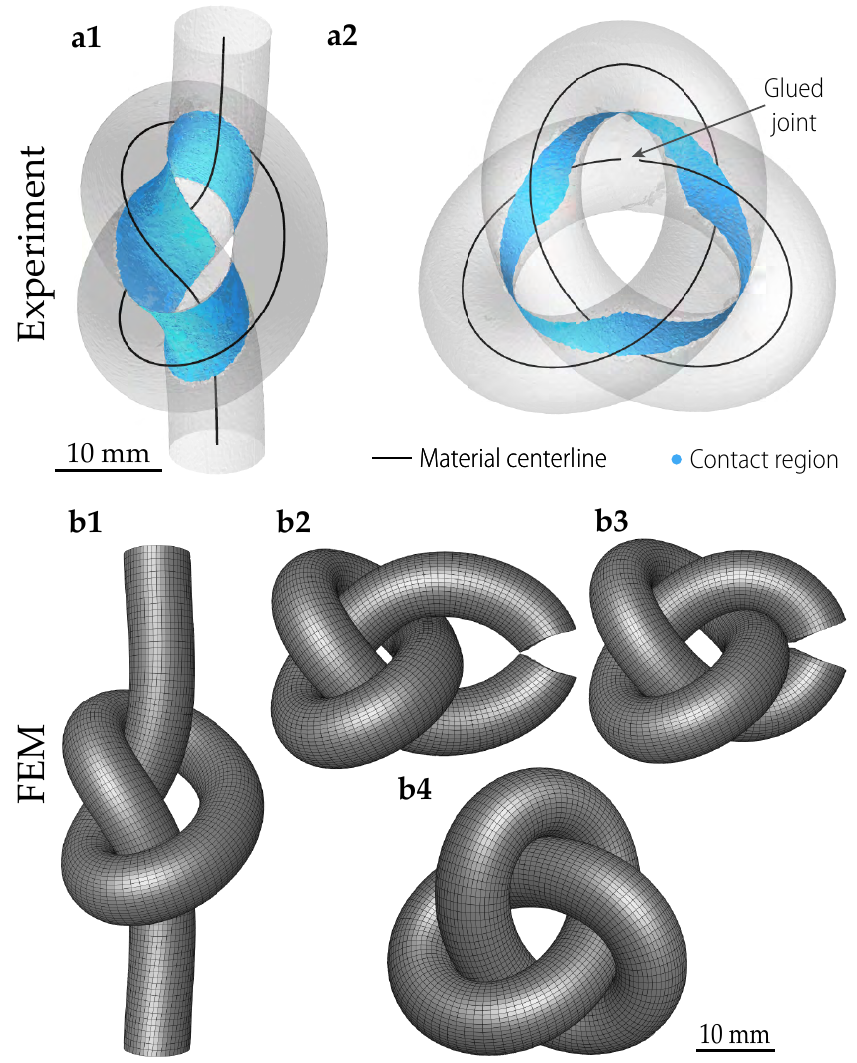}
        \caption{\textbf{Methods to realize and analyze the elastic \textit{open} and \textit{closed} trefoil knots.} \textbf{a1},~Rendering of the reconstructed $\mu$CT-data of an experimental open trefoil knot with a normalized knot length $\Lambda_\textrm{OC}=85.9/D_0=10.1$.
        \textbf{a2},~Rendering of the reconstructed $\mu$CT-data of the elastic closed trefoil knot with a length-to-diameter ratio of $L_0/D_0 = 16.37$.
        \textbf{b1},~FEM computed equilibrium shape of an open trefoil knot.
        \textbf{b2-b4},~Successive rod-end displacements to obtain the FEM closed configuration starting from a tight open trefoil knot.
        }
        \label{Fig2}
\end{figure}

\noindent \textit{2.1.4. Post-processing of $\mu$CT images:}
We quantified the 3D geometry of the knotted rods using $\mu$CT imaging ($\mu$CT100, Scanco Medical), with spacial resolutions (voxel size) of $24.6\,\mu \text{m}$ or $16.4\,\mu \text{m}$ for the open or closed knot configurations, respectively.  An adaptation of the algorithm developed by Grandgeorge and~Baek~\textit{et al.}~\cite{Grandgeorge2020} was used for subsequent post-processing of the tomographic images. In this process, the segmentation of the images leveraged the density difference between the various rod features. The embedded concentric physical fiber allowed us to extract a discrete set of the locations of the centerline coordinates, $\mathbf{r}(s_i)$. The integer~$i$ corresponds to the index of the centerline locations with~$1\le i \le N$ (where $N$ is the total number of centerline points). The application of a Gaussian-weighted moving average filter to $\bm{r}(s_i)$ was necessary (see \secref{Appendix:NoiseReduction}) to compute the discretized curvature of the rod centerline, as described next. 
We first constructed the discrete set of tangent vectors~$\mathbf{e}$ at $s = s_i$ such that $\mathbf{e}(s_i) = \mathbf{r}(s_i + \delta s)-\mathbf{r}(s_i)$, with the increment $\delta s\equiv\| \mathbf{r}(s_{i+1}) -  \mathbf{r}(s_{i}) \|$. The discretization increment was fixed to~$\delta s = 150\,\mu{\rm m}$ and~$\delta s = 100\,\mu{\rm m}$ for the open and closed knot configurations, respectively. We then computed the curvature profiles of the discrete framed curves according to Bergou \emph{et al.}~\cite{Bergou2008}:
\begin{equation}
    K(s_i) = \frac{|2 \mathbf{e}(s_{i-1}) \times \mathbf{e}(s_{i})|}{|\mathbf{e}(s_{i-1})| |\mathbf{e}(s_{i})| + \mathbf{e}(s_{i-1}) \cdot \mathbf{e}(s_{i})}\cdot \frac{1}{|\mathscr{D}(s_i)|},
\end{equation}
with the length of the Voronoi region $|\mathscr{D}(s_i)| = (|\mathbf{e}(s_{i-1})| + |\mathbf{e}(s_{i})|) / 2$. Finally, the regions in the $\mu$CT images corresponding to the thin uniform outer coating layer of Solaris were segmented to reveal the regions of self-contact. The individual contact points on the rod surface are shown in \figref{Fig2}a1~and~a2.

\subsection{Finite element simulations}
\label{Trefoil_FEM}

\begin{figure*}[ht!]
        \centering
        \includegraphics[width=0.82\textwidth]{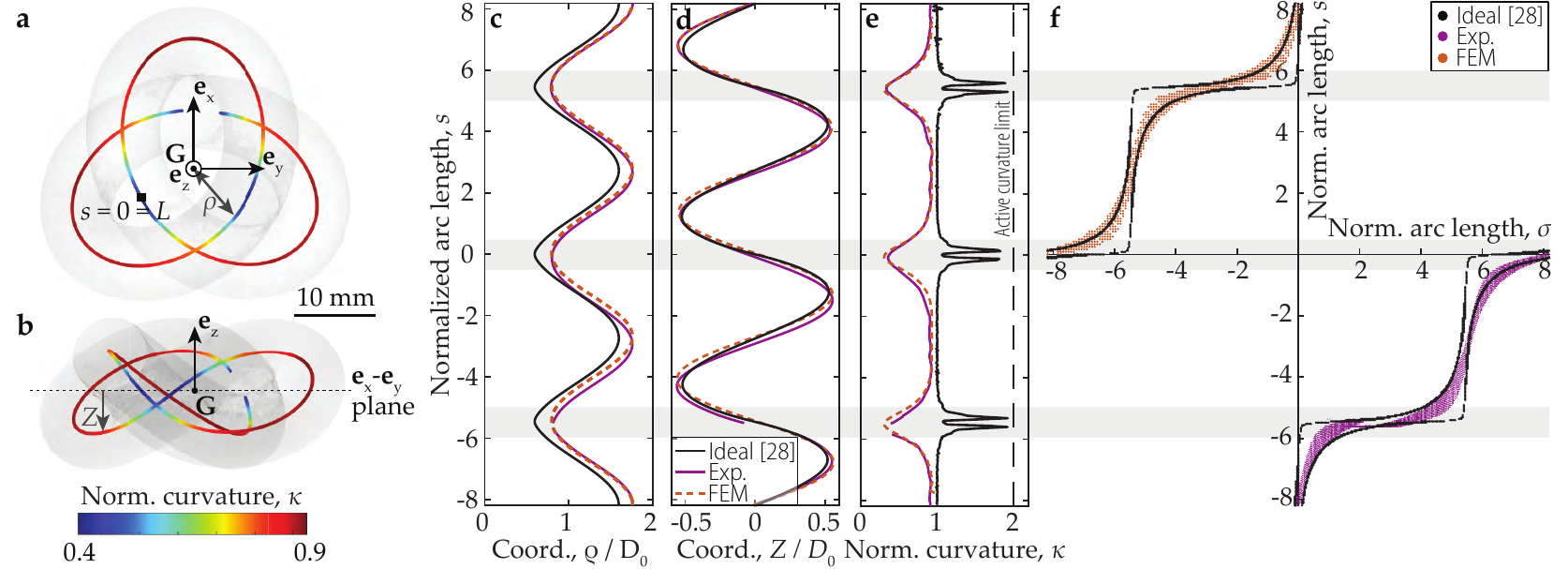}
        \caption{\textbf{Ideal versus elastic closed trefoil knots.}
        \textbf{a}, \textbf{b}~Top and side views of the 3D reconstruction of the experimental closed trefoil knot,
        including the barycenter of the closed centerline curve, $\mathbf{G}$, and the radial and vertical centerline coordinates, $\rho$ and $Z$ respectively. The normalized centerline curvature, $\kappa(s)=KD_0$, is represented by a color-map on the centerline curve.
        \textbf{c},~Comparison of the normalized radial centerline coordinate $\rho/D_0 \equiv \sqrt{X^2 + Y^2}/D_0$. The three arc lengths $S$ are individually normalized and rescaled such that $s=S/L \times L_0/D_0$.
        The shaded areas indicate the inner segments. \textbf{d},~Comparison of the normalized vertical centerline position~$z=Z/D_0$. 
        \textbf{e},~Normalized curvature profile for the ideal and elastic case, including the active curvature limit at $\kappa = K D_0 = 2$. The black solid line (ideal) data are reproduced from Ref.~\cite{Carlen2005, Gerlach2010}.
        \textbf{f},~Contact map showing the characteristic double contact in the ideal case, and the filled area of the equivalent elastic case (experiment and FEM).
        }
        \label{Fig3}
    \end{figure*}

We used the finite element method (FEM, ABAQUS STANDARD 6.14-1, Simulia, Dassault Systems 2014) to simulate the tying of the same knots realized in the experiments. These experimentally validated simulations yield information that cannot be accessed directly through experiment; \textit{e.g.}, the pressure field in the regions of self-contact. Contrariwise, the close agreement between the two (see results in Figs.~\ref{Fig3}, \ref{Fig4}, and \ref{Fig5}) serves as a verification that the experimental configurations are, indeed, the equilibrium ones, with no additional significant experimental artifact.

The FEM computations were performed using the procedure reported recently by Baek~\textit{et al.}~\cite{Baek2020}, involving a dynamic-implicit analysis to capture the geometrically nonlinear deformation of the closed trefoil knot. A rod of diameter $D_0$ and length $L_0$ (paralleling the rod dimensions in the experiments) was meshed using 3D  solid elements with reduced integration (\texttt{C3D8RH}). The number of elements per cross-sectional area was 120 and 190 for the open and closed knots, respectively. The mesh size along the axial direction of the rod was chosen such that the aspect ratio of the elements was close to unity. We modeled the elastomer as an incompressible neo-Hookean material of Young's modulus $E=1.25\,\text{MPa}$. Self-contact of the rod was enforced using a penalty normal force model combined with there being no tangential force (frictionless contact). 

Starting from the initially straight configuration of a rod, we obtained a final knotted geometry by applying a sequence of displacement steps at each extremity of the rod. Firstly, we established a configuration of the open trefoil knot based on the knot-tying procedure described in Baek~\textit{et al.}~\cite{Baek2020} (see \figref{Fig2}b1). Then, we gradually brought the extremities of the rods in contact to establish the closed configuration (see Figs.~\ref{Fig2}b2-b3), with the final equilibrium configuration of the closed trefoil knot presented in \figref{Fig2}b4. The two extremities were constrained using the ABAQUS command \texttt{*COUPLING}, which enables the extremities to be displaced while allowing their cross-section to deform. Throughout, we ensured that the simulation was quasi-static. 


\section{Ideal versus elastic \textit{closed} trefoil knots}
\label{cTrefoil_Results}

Having described our experimental and numerical toolbox, we proceed by quantifying the similarities and dissimilarities between physical and ideal \textit{closed} trefoil knots, with the analogous discussion of the \textit{open} case appearing in the next section. A closed knot offers the advantage of having a closed centerline curve with matching periodic boundary conditions; its configuration is not subject to external factors such as applied external forces. As the experimental material is elastic, a trefoil knot can be tied in tubes with a wide range of aspect ratios of $L_0/D_0$ of undeformed centerline length to undeformed cross-section diameter. Cases with $L_0/D_0$ large would correspond to loose knots as considered in~\cite{Audoly2007,Clauvelin2009,Jawed2015}. Cases with $L_0/D_0$ small would require large extension just to be able to close the centerline to form the knot, with associated large tensions, and presumably associated large cross-sectional deformation. A systematic study of dependence on a range of chosen values for $L_0/D_0$ is beyond the scope of the current work. Instead we chose the single critical value $L_0/D_0 = 16.37$, which is a good approximation to the smallest value known to be possible in the ideal geometric theory with an inextensible centerline and undeformable cross-section (and no bending stiffness). We would expect the resulting experimental equilibrium configuration to be relatively tight, and with relatively small centerline extension and cross-sectional deformation. 
After a closed trefoil is tied on the physical elastic rod (with undeformed rod length~$L_0=139.1\,\text{mm}$ and cross-section diameter $D_0=8.5\,\text{mm}$), the observed stretch of its centerline is 1.070 in experiment and 1.082 in FEM-simulation. The overall length-to-diameter ratio of the stretched rod, $L/D$, was measured to be 18.12 and 18.53 in experiments and FEM-simulations, respectively. Note that we define the average reduced diameter as $D = D_0[1-\nu(L-L_0)/L_0]$, where  the Poisson's ratio is $\nu \approx 0.5$.
To further compare our results with those of the ideal geometric theory, we take the observed small axial strain into account by using the normalized and rescaled arc length $s=S/L \times L_0/D_0$, with the stretched rod length, $L$, and the ideal normalized rod length $L_0/D_0 = 16.37$, while also assuming that the axial strain is constant along $S$. 

To perform a comparison between the \textit{centerline coordinates} $\mathbf{r}(s)=(X(s),\,Y(s),\,Z(s))$ of the elastic and the ideal closed trefoil knots, we introduce (following Ref.~\cite{Carlen2005}) cylindrical coordinates in the Cartesian basis $\{\mathbf{e}_{x},\mathbf{e}_{y},\mathbf{e}_{z}\}$, as shown in \figref{Fig3}a,b. The knot lies flat on the $\mathbf{e}_x\text{-}\mathbf{e}_y$-plane, and the origin is chosen by the condition that the center of mass, or barycenter, of the centerline curve $\mathbf{G} = \sum_{i = 1}^{N}{\mathbf{r}(s_i) \delta s_i} / L$ lies on the $\mathbf{e}_{z}$-axis. (Here, $N$ is the number of discretization points, $L$ is the stretched rod arc length, and~$\delta s_i$ is the length of the $i^\text{th}$ segment between two successive discretized centerline points.) In \figref{Fig3}c, we compare the radial distance between the centerline and the barycenter axis, quantified as $\rho(s)\equiv \sqrt{X^2(s) + Y^2(s)}$, for the experimental, FEM and ideal knot cases (with three individually scaled arc lengths on ordinate, but all plots with the same common length scale on abscissa). The experimental and FEM data are in excellent agreement. Compared to the ideal knot, the experimental and FEM closed knots exhibit a radial inflation, presumably due to elasticity effects, as evidenced by the horizontal offset of the $\rho$ data. 
For example, $\rho/D_0$ differs by $0.20$ and $0.16$ in the inner segments (minima; shaded) and the outer segments (maxima of $\rho/D_0$ curves), respectively.
Moreover, the effect of the cross-sectional deformation is reflected in the amplitude of $\rho$ for the elastic knot which is $0.95\,D_0$ ($D_0$ for the ideal case).  
To complete the comparison of the cylindrical coordinates, in \figref{Fig3}d, we present the rescaled vertical centerline coordinate, $Z(s)/D_0$, for the three cases, which, interestingly, shows an excellent match between the ideal and the elastic closed knots, unlike the $\rho$ data presented in \figref{Fig3}c.

Based on the $\mu$CT and FEM data, we construct a two-dimensional contact map; the projection of the contact surface onto the arc length $s$ vs.\ arc length $\sigma$ plane. To assemble this contact map, each point in the contact surface is assigned to the two closest centerline positions of the knotted rod, at arc lengths $s$ and $\sigma$.

In Fig~\ref{Fig3}f, we plot the contact map for the ideal case~\cite{Carlen2010} (black solid lines), together with the corresponding data extracted from $\mu$CT and FEM. Note that by construction the arc length contact map is point-symmetric with respect to $s=\sigma=0$~\cite{Carlen2010}. Consequently, due to this symmetry, we only present one half of the $\mu$CT and FEM contact data, respectively in the lower-right and upper-left quadrants of the $s-\sigma$ plot in Fig~\ref{Fig3}e. We observe that, whereas, for the ideal knot, there are precisely two contact points $\sigma_1$, $\sigma_2$ for each $s$ value, the physical knots exhibit an extended contact region with a range of $\sigma$ values for each $s$ value. Moreover, we find that the contact set for the physical knot is a surface that lies fully inside the double contact lines (black lines) of the ideal closed trefoil knot; the geometric model acts as an outer skeleton for the elastic case.  This filled (areal) contact region for the physical case, replaces the double-line contact in the ideal knot (see \secref{Introduction} and ~\ref{Appendix:Ideal_cTrefoil}) due to cross-sectional deformation. The mismatch between the ideal and the elastic cases is particularly evident in the inner segments; there, the corners of the geometric contact set are not filled in the elastic case. 

In \figref{Fig3}e, we plot the \textit{curvature profiles} of the elastic and the ideal trefoil knots. The curvature data are also presented in \figref{Fig3}a (see color-map), along the centerline of the experimental case. The elastic knot exhibits plateaus in the three outer segments with average normalized curvatures of $\kappa = K D_0 \approx 0.93$ (whereas $\kappa \approx 1.00$ for the ideal knot). Despite these close values in the outer segments, the behavior in the inner segments is strikingly different between the elastic and ideal cases; the elastic knot exhibits clear curvature minima, whereas the ideal model predicts twin curvature peaks approaching the active curvature limit  $\kappa = 2$, separated by a local minimum~\cite{Carlen2005}. We hypothesize that this difference in curvatures between the two cases is rooted in the cross-section deformations allowed in 3D elasticity, which we address further in the next section, in the context of the \textit{open} trefoil knot.


\section{Ideal versus elastic \textit{open} trefoil knots}
\label{oTrefoil_Results}

The \textit{open} trefoil knot allows us to directly control the level of tightness by applying forces to the rod extremities, to study the role of elasticity more systematically. This feature is not possible in the closed case since the extremities are, naturally, `\textit{glued}' together. We will employ the experimental and numerical toolbox that we developed for the \textit{closed} trefoil knot to explore the similarities and dissimilarities between the elastic open trefoil knot and the corresponding ideal case~\cite{Pieranski2001,Pieranski2001a}.

\begin{figure}[ht!]
        \centering
        \includegraphics[width=0.8\columnwidth]{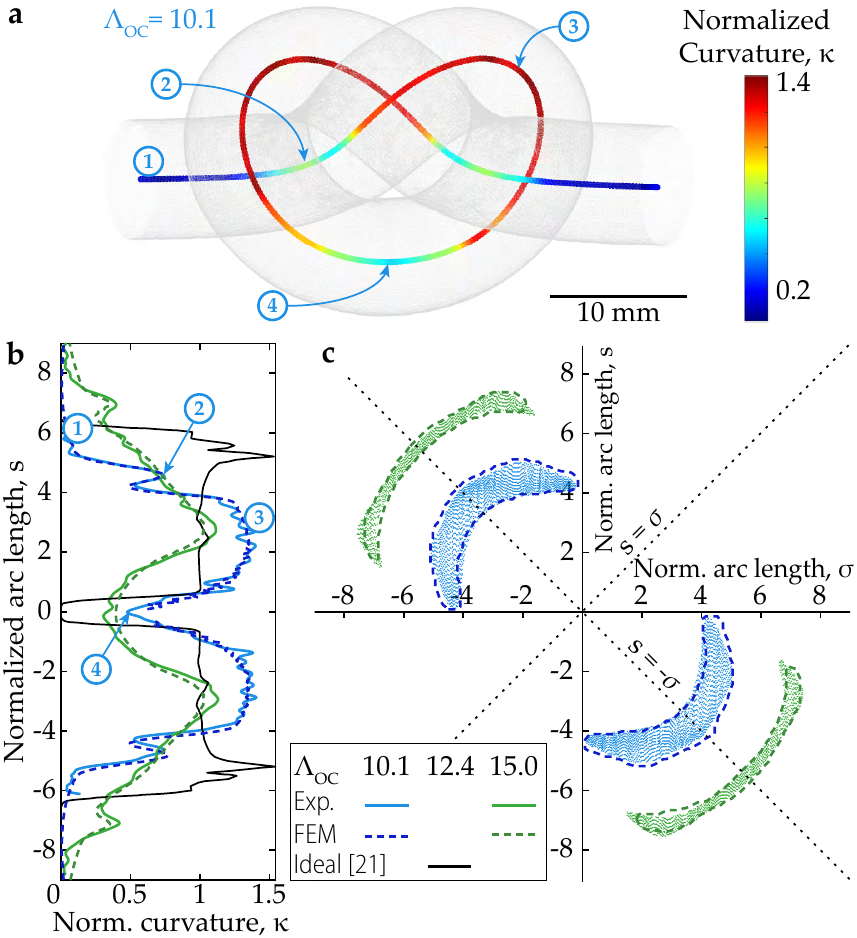}
        \caption{\textbf{Ideal versus elastic open trefoil knots.}
        \textbf{a},~Reconstruction of an experimental open trefoil knot with $\Lambda_\textrm{OC}=10.1$. The centerline of the knot is overlaid by the color-map of the normalized curvature profile $\kappa(s)=KD_0$. The following features are referred to in the text: (1)~knot entrance, (2)~inner segment, (3)~outer segment, and (4)~central loop.
        \textbf{b},~Normalized curvature profiles. Experimental (solid colored lines) and numerical data (dashed colored lines) for two normalized knot lengths~$\Lambda_\textrm{OC}$=\{15;\,10.1\} compared to the geometric description according to Piera{\'{n}}ski~\textit{et al.}~\cite{Pieranski2001} (thin black line). The arc length $S$ is normalized such that $s=S/D_0$.
        \textbf{c},~Contact regions mapped into the $s-\sigma$ space:  $\mu$CT data (filled area) and FEM data (dashed lines, only the region boundaries are shown). The normalization of the arc length $S$ is $s=S/D_0$ and $\sigma=S/D_0$.}
        \label{Fig4}
    \end{figure}

In \figref{Fig4}a, we present the 3D reconstruction of an experimental open trefoil knot (normalized knot length of $\Lambda_\textrm{OC} = 10.1$), with the measured normalized curvature profile, $\kappa(s)=KD_0$ superposed onto the centerline. This curvature profile is qualitatively similar to what we observed in \secref{cTrefoil_Results} for the physical, closed trefoils, with minima at the inner segments (region (2) in \figref{Fig4}a). In \figref{Fig4}b, we plot the experimental and FEM-computed $\kappa(s)$ profiles for the two elastic knots that we investigated, with normalized knot lengths of $\Lambda_\textrm{OC} = 10.1$ and $15.0$. By way of example, we describe the physical knot with $\Lambda_\textrm{OC} = 10.1$, referring to the features labeled in \figref{Fig4}a and b while traveling along  arc length (increasing $s$). Soon after the knot entrance~(1), the vanishing curvature of the almost straight elastic rod rises to a local maximum, in the central region of the inner segment~(2). The transition of the rod from the inner to the outer segment has a curvature drop, followed by an abrupt rise. The normalized curvature then reaches its maximum value in the outer segment~(3). In this high-curvature region, we find that $\kappa > 1$ over a wide range of $s$ due to cross-sectional deformation of the elastic rod. Eventually, there is a local curvature minimum at the central part of the loop~(4). 

The curvature profile of the ideal open trefoil knot in its tightest configuration ($\Lambda_\textrm{OC} = 12.4$) obtained by Pieranski~\textit{et al.}~\cite{Pieranski2001} is also shown in \figref{Fig4}b, superposed onto the elastic profiles for comparison. There are important qualitative differences between the ideal and elastic results. For example the prominent curvature peaks occur at different locations and with different shapes between the two cases, a difference that can be attributed to elastic deformation of the cross-sections and the centerline. 

In \figref{Fig4}c, we map the contact region for elastic knots with $\Lambda_\textrm{OC} = 10.1$ and $15.0$ (blue and green regions, respectively), extracted from the $\mu$CT and FEM data.  For the experiments, the full contact region in the $s-\sigma$ space is plotted, whereas, to aid comparison, only the outer boundaries of the contact regions are shown for the FEM data (dashed lines). Again, FEM and experiments are in excellent agreement. Naturally, the contact map of a self-contacting rod is  symmetric with respect to the $s=\sigma$~axis. Indeed, if contact occurs at the centerline arc length $s=a$ with the arc length $\sigma=b$, then it also occurs at $s=b$ with $\sigma=a$. Moreover, the symmetric nature of the overhand knot about $s=0$ introduces the axis of symmetry $s=-\sigma$ on its contact map.

  \begin{figure*}[ht!]
        \centering
        \includegraphics[width=0.8\textwidth]{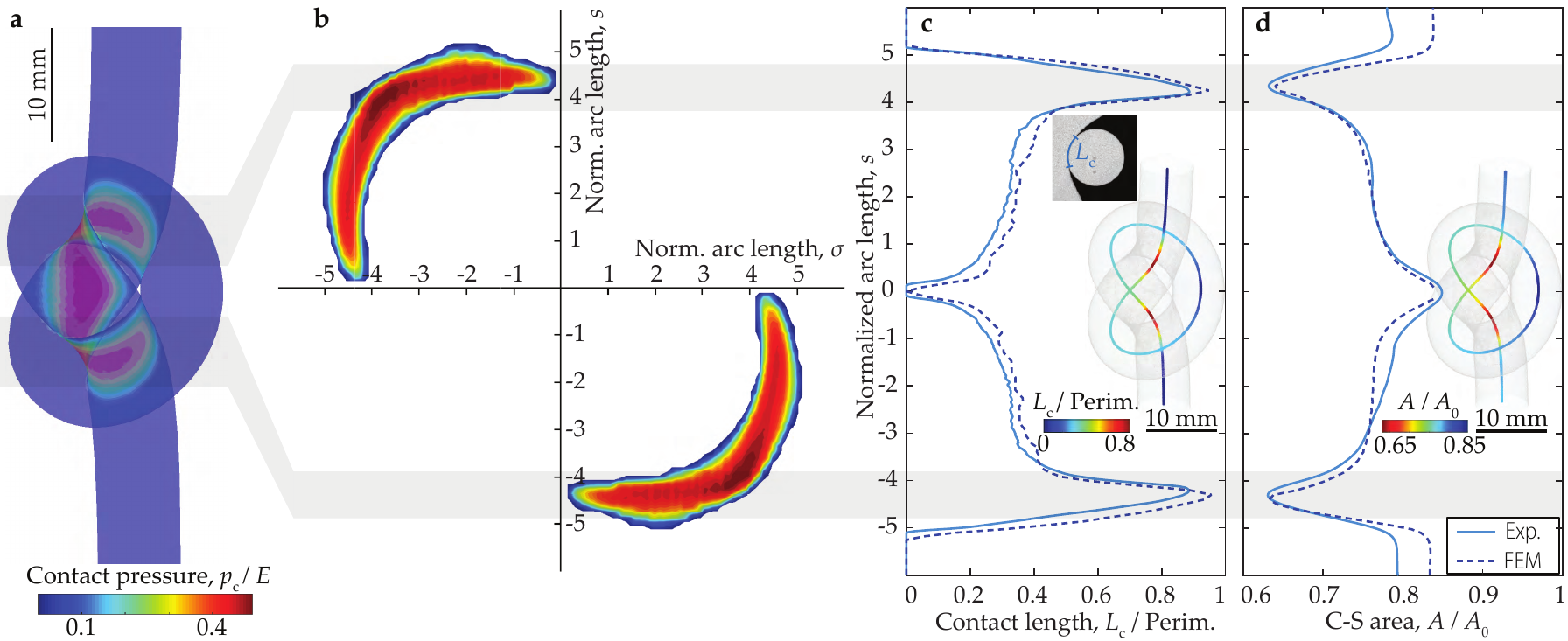}
        \caption{\textbf{Constriction at the entrance and exit of elastic tight open trefoil knots.} \textbf{a},~Numerical FEM contact pressure on the 3D knot of the tight configuration ($\Lambda_\textrm{OC} = 10.1$). The shaded regions indicate the inner segments at the knot entrance/exit. \textbf{b},~Numerical contact pressure map for tight configuration ($\Lambda_\textrm{OC} = 10.1$). The normalization of the arc length $S$ is $s=S/D_0$ and $\sigma=S/D_0$. \textbf{c},~Circumferential ($\mu$CT-~and FEM-data) contact set width  $L_\textrm{c}$ along the arc length, showing clear peaks at the entrance/exit of the knot.
        \textbf{d},~Cross-sectional area ($\mu$CT-~and FEM-data) along the arc length, normalized by the cross-sectional area of the undeformed rod. The quantities plotted both in \textbf{c} and \textbf{d} are also represented on the respective insets, using a color-coded centerline of the 3D reconstruction of the knot.
        }
        \label{Fig5}
    \end{figure*}

From the simulations, we extracted data for the contact pressure (normal traction) at the regions of self-contact. In \figref{Fig5}a, we present a snapshot of the elastic knot with $\Lambda_\textrm{OC}=10.1$, including the contact regions onto which we superpose the contact pressure
(normalized by the Young's modulus~$E$). The contact pressure map is shown in \figref{Fig5}b, using a similar representation (in the $s-\sigma$ space) used in \figref{Fig4}c for the contact map. The highest contact pressure is found along the entire central region of the contact set, with maximum characteristic normalized values of $p / E \approx 0.44$. Note that the knot entrance/exit (shaded regions in \figref{Fig5}b) correspond to regions of localized pressure, aligned perpendicularly to the rod centerline. To further quantify the localization of deformation along the knot, in \figref{Fig5}c, we present measurements of the circumferential contact set width profile $L_{c}(s)$, normalized by the total perimeter of a rod cross-section at arc length~$s$. We observe sharp peaks of $L_{\textrm{c}}$ at the inner segments ($-4.8 \lesssim s\lesssim -3.8$ and $3.8 \lesssim s \lesssim 4.8$), where up to 90\% of the circumference of the cross-section is in self-contact. The regions of pronounced contact pressure (\figref{Fig5}b) in combination with the sharp circumferential contact width peaks (\figref{Fig5}c) lead to localization of high contact pressure in a narrow region with a small range of arc lengths. Consequently, as shown in \figref{Fig5}d, where we quantify the profile of deformed cross-sectional area as a function along the centerline of the rod, the cross-section of the inner rod segment is elastically constricted by up to $\sim63\%$ compared to its rest cross-section area; such localized constrictions in knots are typically referred to as \textit{nip} regions~\cite{Ashley1944}.

\section{Discussion and Conclusions}
\label{Conclusions}

We have systematically quantified the shapes of {\it physical} trefoil knots, in both \textit{closed} and \textit{open} configurations. Excellent agreement was found in all considered quantities between FEM and experiment. For the latter, we made extensive use of X-ray micro-computed tomography, gaining access to volumetric information, including centerline curvature and cross-sectional deformation profiles. In parallel, the experimentally validated FEM enabled us to quantify the contact pressure field, which is not available in experiment. Direct comparisons were also established between the experimental and FEM data for elastic trefoil knots and prior numerical computations of their (purely geometric) ideal shape counterparts.

For both open and closed physical trefoil knots the contact sets observed in both experiment and FEM were 
\textit{smooth surfaces}, with a positive contact set width $L_\textrm{c}$, \textit{i.e.}\ finite strips. For the closed trefoil, the physical contact surface is actually a closed strip, which, as an additional topological observation, we remark is a one and a half turn M\"obius band (the more common M\"obius band has only a single half turn) and so is non-orientable (it has only one face) and only one edge. Moreover, for such 1.5-turn M\"obius bands the single edge itself forms a trefoil knot (see Supplementary Movie~1). This is perhaps at first sight surprising, but the topology of the contact strip is inherited from the topology of the contact line of the ideal closed trefoil configuration, where it is already understood that the contact set in 3D is a closed curve that is itself a trefoil knot~\cite{Carlen2005}. Just as for the 2D arc length contact sets (cf.\ \figref{Fig3}), where the 2D contact region of the elastic configuration fills the outer skeleton provided by the double-contact line of the ideal geometric model when elastic deformation of the cross-section is allowed, the 3D ideal contact set curve acts as skeleton, which is fattened, or bridged, to arrive at a 3D physical contact surface strip, whose topology is inherited from the ideal case.

In the comparison between the elastic and ideal cases of trefoil knots, we found that their curvature profiles were not just quantitatively different, but also qualitatively different. In both open and closed cases, elasticity regularizes the curvature peaks within the inner segment that are predicted by the purely geometric model. To gain insight into the discrepancies between the elastic and the ideal systems, we focused on the open configuration, allowing us to systematically vary the knot tightness. The curvature peaks of the elastic system occur in the outer segment, for both looser and tight knots, contrary to the geometric counterpart, where they appear at the knot's entrance/exit regions. The contact pressure distribution extracted from FEM exhibited  localized regions at the entrance of the knot (inner segments). This pressure localization leads to a prominent cross-sectional deformation in the inner segments, acting as local constrictions in these nip regions. 

As reported by C.W.~Ashley in his comprehensive reference manual on knots~\cite{Ashley1944}, ``\textit{a rope is weakest just outside of the entrance of the knot}''; a finding that is commonly confirmed by practical experience in knotted filaments. The significant reduction in the cross-sectional area reported in Fig~\ref{Fig5}d at the entrance/exit of tight elastic open knots could act as a precursor for weak spots on knotted filaments. Our interpretation is different from that of Piera{\'{n}}ski \textit{et al.}~\cite{Pieranski2001a}, who attributed the onset of failure to regions of high centerline curvature, computed using their purely geometry model, which our results demonstrate to be in strong disagreement with the curvature profile of physical knots. Our investigation highlights that a mechanics-based approach, going beyond pure geometry, will be necessary to rationalize knot failure. Given the high level of tightening in functional knots tied onto elasto-plastic material filaments, these constriction regions are prone to local plastic deformation~\cite{Uehara2007}. The effect of plasticity on the equilibrium shape of physical knots remains an open question, which we hope to untangle in future studies.\\

\noindent \textbf{Declaration of Competing Interest.}~The authors declare that they have no known competing financial interests or personal relationships that could have appeared to influence the work reported in this paper.\\

\noindent \textbf{Acknowledgments.}~The authors thank A. Flynn for fruitful discussions.
This work was supported by the Fonds National de la Recherche, Luxembourg (12439430), the Grants-in-Aid for JSPS Overseas Research Fellowship (2019-60059), and by the Swiss National Science Foundation (Award 200020-18218 to JHM).

    \bibliographystyle{elsarticle-num-names}
    \bibliography{ProjectKnots}

\appendix

\section{A brief overview of ideal closed knots}

In the Introduction section of the main text, we defined tightest or ideal knot shapes as the centerline configuration with the shortest possible length amongst all those tied in a closed loop of an idealized rope, which is taken to mean a filament with an undeformable circular cross-section of prescribed diameter, and inextensible centerline (and vanishing bending stiffness so the problem has no mechanics, only geometry). In this Appendix, for completeness, we provide a brief overview of existing literature on ideal knots (primarily from the Mathematics community, but which we hope may be of some interest to the Mechanics community). Specifically, we focus on a more technical description of the necessary conditions that must be satisfied by ideal shape centerline curves, and the double-contact feature that is manifested in the ideal trefoil knot and other geometries (not-knots).

\subsection{Ideal shapes}
\label{Appendix:Ideal_shapes}

Ideal shapes are known to exist for all standard knot types~\cite{Gonzalez2002a} with centerlines that are $C^{1,1}$~curves, meaning that the centerline has a continuously varying unit tangent at every point, and a curvature that is defined \textit{almost} everywhere, but \textit{not} everywhere. In particular the curvature can be discontinuous. For example, a straight line segment joined to an arc of a circle with matching tangents, but a discontinuous curvature, can form part of an ideal shape, and numerics strongly suggest that straight line segments and discontinuities in curvature do arise in ideal shapes, for example on composite knots~\cite{Gonzalez1999}. As a side note, we point out that knowing the fine detail of the precise smoothness or regularity of ideal knot shapes is important in designing good numerical algorithms to approximate them. In addition to the circular centerline of the ideal shape of the unknot, the only other known explicit ideal shapes are comparatively simple, piece-wise planar, ideal shapes for certain \textit{links} (\textit{i.e.}, knots with multi-component centerlines)~\cite{Cantarella2002, Grandgeorge2020}.

The first conditions that must be satisfied by ideal knot shapes were derived in Ref.~\cite{Gonzalez1999}, in terms of the \textit{global radius of curvature}. Technically, these results depended on the slightly too strong assumption of a $C^2$ centerline. Extensions to the weaker and sharp hypothesis of a $C^{1,1}$ centerline were obtained in Ref.~\cite{Schuricht2003}, where the appropriate Euler-Lagrange equations were also related to force balance. The necessary conditions that must be satisfied on an ideal shape include the three-way alternative that every point along an ideal centerline must be either (i) part of a straight segment, or (ii) local curvature must achieve its maximal value $2/D_0$ (as is the case everywhere for the circular ideal shape for the unknot), or (iii) be at one end of a locally minimal distance, or contact chord, of length $D_0$ between two distinct points on the centerline.

\subsection{The ideal closed trefoil knot}
\label{Appendix:Ideal_cTrefoil}

\begin{figure}[h!]
        \centering
        \includegraphics[]{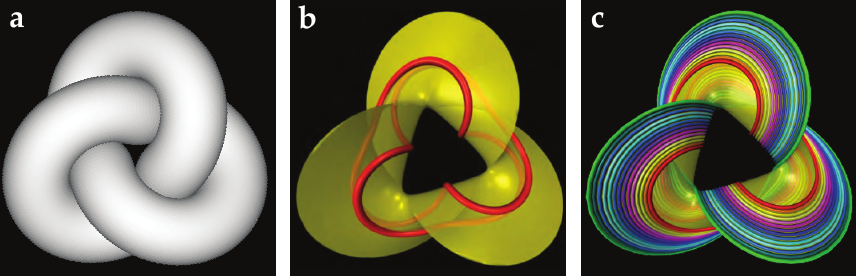}
        \caption{\textbf{3D Visualizations of the ideal closed trefoil knot}. (Adapted with permission from \cite{Gerlach2010, Carlen2010}.)  \textbf{a}, a solid tube visualization, which obscures the inner structure of the one-parameter family of contact chords shown as a translucent yellow surface in panel \textbf{b} (at a slightly larger scale). The sharp edge of the surface is the centerline of the knot. The red curve traced out by the center points of the contact chords is the contact set where the tube of panel \textbf{a} touches itself. The contact curve can be seen to itself be a trefoil knot by the smooth homotopy from the contact curve (red) to knot centerline (green) illustrated in panel \textbf{c}, where each of the non-intersecting multi-colored closed curves lies on the yellow contact surface.}
        \label{Fig6_Appendix}
\end{figure}

As mentioned in the main text, we believe that the most accurate numerical approximation to the ideal closed trefoil currently available is that provided by Przybyl~\textit{et al.}~\cite{Przybyl2014}, with $L_0/D_0 = 16.371476 \dots $. This computed value of $L_0/D_0$ is a rigorous (to machine arithmetic precision) upper bound to the actual ideal value, and very probably the upper bound is rather close to the actual, unknown ideal value. However, rather than comparing many digits of accuracy in the ideal value of $L_0/D_0$, the present study seeks to compare \textit{features} of computed ideal closed trefoil shapes with both experiment and FEM simulation, which include a combination of the elastic effects of bending and deformation of cross-sections. The features that we compare were first described for the ideal geometric trefoil by Carlen~\textit{et al.}~\cite{Carlen2005},  and subsequently confirmed and better visualized on improved simulation data, as fully described in \cite{Gerlach2010, Carlen2010}, from where the images in Fig.~\ref{Fig6_Appendix} are reproduced.

For the ideal closed trefoil, visualized as a solid tube in Fig.~\ref{Fig6_Appendix}a, each point along the tube centerline is in fact at the end of \textit{two} distinct contact chords. This gives rise to the double contact lines in the $(s,\sigma)$ plane shown in Fig.~\ref{Fig3}f in the main text. In addition the maximum local curvature of $2/D_0$ is very close to being attained at six points, as shown in the spikes in Fig.~\ref{Fig3}e in the main text. Furthermore, the curvature is nowhere close to vanishing, so that no straight segment arises in the ideal trefoil knot centerline. The double-contact feature is present along the full arc length of the trefoil knot, which means that there is a one-parameter family of double contact chords which trace out a surface in 3D, as shown in Fig.~\ref{Fig6_Appendix}b, where the sharp edge of the translucent yellow surface is the centerline  curve of the ideal trefoil shape.  The 3D contact set for the ideal trefoil is a closed curve lying on the surface of the tubes visualized in Fig.~\ref{Fig6_Appendix}a, but is obscured. This contact line is also traced out by the mid-points of the contact chords (red curve in Fig.~\ref{Fig6_Appendix}b). The contact curve can be seen to itself be a trefoil knot by the homotopy illustrated in Fig.~\ref{Fig6_Appendix}c, where there is a family of non-intersecting multi-colored closed curves that deform along the contact surface from the contact curve (red) to the knot centerline (green).

Analogous double-contact phenomena have previously been reported for infinite double helices, depending on the pitch angle~\cite{Stasiak2000,Gonzalez2002}, and in an ideal orthogonal clasp problem~\cite{Starostin2003}. Maritan~\textit{et al.}~\cite{Maritan2000,Stasiak2000,Gonzalez2002} also showed that in an optimal packing problem, single helices frequently arise with both double contact chords and maximal curvature, and that the associated critical aspect ratio of this special helix arises for the $C_\alpha$ carbons in $\alpha$ helical segments of protein crystal structures. Thus the observed phenomena of double contact chords with additionally maximal curvature, is perhaps not as exceptional as it might first appear. 


\section{Smoothing of the raw data to reduce `noise' in the curvature computation}
\label{Appendix:NoiseReduction}

\begin{figure}[h!]
        \centering
        \includegraphics[]{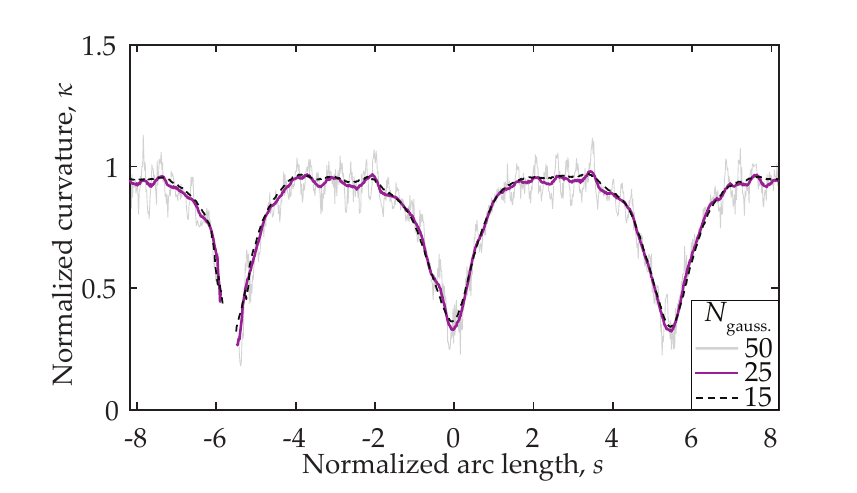}
        \caption{\textbf{Test for the smoothing of the curvature computed from the centerline data for the experimental closed trefoil knot.} A Gaussian-weighted moving average filter with changing size of the smoothing window allowed to find the trade-off value between noisy and over-smoothed curves. The selected value for the window size used for the data presented in the main text is $N_\textrm{gauss}=25$ (\textit{i.e.}, $\sigma = 39$).}
        \label{Fig7_Appendix}
\end{figure}

As described in the main text (Sec.~2.1.4), we computed the curvature profiles of the rod centerline by the numerical differentiation of ${\bm r}(s)$. Prior to this differentiation, we applied a Gaussian-weighted moving average filter (command \texttt{smoothdata} in Matlab 2019) to ${\bm r}(s)$, with a window size defined by $\sigma = \texttt{round}(N_\textrm{b}/N_\textrm{gauss})$. To test the fidelity of the computed curvature data, given the discrete nature of the raw data, we performed a parametric test of the filter on the closed trefoil knot with $L_0/D_0 = 16.37$ (the same configuration studied experimentally in the main text). In this test, we fixed the total number of discrete centerline points $N_\textrm{b}=984$ for the closed trefoil knot, and systematically varied $N_\textrm{gauss}=\{15,\, 25,\, 50\}$. Without the filter (\textit{i.e.}, $\sigma=1$, corresponding $N_\textrm{gauss}=984$), the data would be far too noisy for analysis. In Fig.~\ref{Fig7_Appendix}, we present profiles for normalized curvatures, $\kappa(s)$, for decreasing values of $N_\textrm{gauss}$ (the data is increasingly smoothed as $N_\textrm{gauss}$ decreases). We selected the window size of $\sigma = 39$ (\textit{i.e.}, $N_\textrm{gauss}=25$), which reasonably suppresses noise while not over-smoothing the curvature features.

\section{Description of Supplementary Movie 1}
\label{Appendix:supp_movie_1}

\noindent \textbf{Homotopy between the centerline of the physical closed trefoil knot and the rim of its contact shape.} Motivated by the known homotopy between the contact line and the knot centerline for numerically simulated, geometrically ideal, configurations of the closed trefoil knot (cf.\ Fig.\ \ref{Fig6_Appendix}) we provide an animation of a three-dimensional rendering constructed from the $\mu$CT data for the physical closed trefoil knot presented in Fig.~1a1, Fig.~2a2, and Fig.~3a,b of the main text. The knotted elastomeric rod (VPS32) has a rest length $L_0 = 139.1$~mm and rest diameter $D_0 = 8.5$~mm ($D_0/L_0 = 16.37$). After locating the centerline of the knotted rod (black curve), as well as its self-contacting surface (blue surface), we extract the rim of the contact surface (red curve). Without undergoing self-crossings (fixed topology), the rim of the contact surface is smoothly morphed into the rod centerline, thus revealing the homotopy between the rim of the contact shape and the rod centerline. To perform this morphing, we first parametrized the rim of the contact shape as $\mathbf{R}(s^*)$, where $s^*$ is the arc length along the rim (of total length $L_R$). The intermediate morphing curve, $\mathbf{w}(s)$, ranges from the rim curve to the centerline curve (parametrized as $\mathbf{r}(s)$), following the parametrized deformation $\mathbf{w}(s) = (1-t) \, \mathbf{R} \,(s \, L_R/L) + t \,\mathbf{r}(s)$, where $L$ is the total length of the centerline curve, and $0 \le t \le 1$ is the morphing parameter.


\end{document}